\documentclass[pra,twocolumn,showpacs,preprintnumbers,amsmath,amssymb,%
nofootinbib,floatfix]{revtex4}
\usepackage{graphicx}
\usepackage{dcolumn}
\usepackage{bm}
\usepackage{hyperref}
\usepackage{color}
\usepackage{ulem}

\newcommand{\Eq}[1]{Eq.\@ (\ref{#1})}

\newcommand{\Fig}[1]{Fig.\@ \ref{#1}}
\newcommand{\Figs}[1]{Figs.\@ \ref{#1}}
\newcommand{\Ref}[1]{Ref.\@ \cite{#1}}
\newcommand{\Refs}[1]{Refs.\@ \cite{#1}}
\newcommand{\Sec}[1]{Sec.\@ \ref{#1}}

\newcommand{\ave}[1]{\langle #1\rangle}

\newcommand{\vek}[1]{\bm{\mathrm{#1}}}
\newcommand{\nablav}{\bm{\nabla}}
\newcommand{\kv}{\vek{k}}
\newcommand{\pv}{\vek{p}}
\newcommand{\rv}{\vek{r}}

\newcommand{\eq}{\mathit{eq}}
\newcommand{\fbar}{\bar{f}}
\newcommand{\ho}{\mathit{ho}}
\newcommand{\rad}{r}

\newcommand{\trap}{T}
\newcommand{\Vtrap}{V_T}
\newcommand{\TOF}{\mathit{TOF}}
\newcommand{\Vhat}{\hat{V}}

\newcommand{\wbarrad}{\bar{\omega}_\rad}
\newcommand{\wrad}{\omega_\rad}

\newcommand{\loc}{\mathit{loc}}

\newcommand{\pulse}{\mathit{pulse}}
\newcommand{\step}{\mathit{step}}

\newcommand{\calT}{\mathcal{T}}

\DeclareMathOperator{\sgn}{sgn}

\DeclareMathOperator{\Imag}{Im}

\renewcommand{\Im}{\Imag}

\newcommand{\dgamma}{\frac{d^2rd^2p}{(2\pi)^2}}
\newcommand{\dgammatext}{d^2rd^2p/(2\pi)^2}

\begin{document}

\title{Damping of the quadrupole mode in a two-dimensional Fermi gas}
\author{Silvia Chiacchiera}
\affiliation{Centro de F{\'i}sica Computacional, Department of Physics,
  University of Coimbra, P-3004-516 Coimbra, Portugal}
\author{Dany Davesne}
\affiliation{Universit{\'e} de Lyon, F-69622 Lyon, France;
  Univ. Lyon 1, Villeurbanne;
  CNRS/IN2P3, UMR5822, IPNL}
\author{Tilman Enss}
\affiliation{Institut f\"ur Theoretische Physik, Universit\"at Heidelberg,
  D-69120 Heidelberg, Germany}
\author{Michael Urban}
\affiliation{Institut de Physique Nucl{\'e}aire, CNRS/IN2P3 and
  Universit\'e Paris-Sud 11, F-91406 Orsay Cedex, France}
\begin{abstract}
In a recent experiment [E. Vogt et al., Phys. Rev. Lett. {\bf 108},
  070404 (2012)], quadrupole and breathing modes of a two-dimensional
Fermi gas were studied. We model these collective modes by solving the
Boltzmann equation {\it via} the method of phase-space moments up to
fourth order, including in-medium effects on the scattering cross
section. In our analysis, we use a realistic Gaussian potential
deformed by the presence of gravity and magnetic field gradients. We
conclude that the origin of the experimentally observed damping of the
quadrupole mode, especially in the weakly interacting (or even
non-interacting) case, cannot be explained by these mechanisms.
\end{abstract}
\maketitle

\section{\label{sec:intro}Introduction}
Two-dimensional (2D) Fermi systems are particularly interesting, since
both quantum and interaction effects are in this case stronger than in
three dimensions (3D). The first experimental realization of a 2D
Fermi gas with trapped atoms was reported in 2010
\cite{Martiyanov}. The configuration obtained in this experiment and
in the subsequent ones is an array of pancake-shaped clouds, obtained
by slicing a 3D cloud with a one-dimensional periodic potential.
These gases can be considered as 2D ones if the motion of particles in
the axial direction is frozen to the lowest energy level.

In a recent experiment, the collective breathing and quadrupole modes
of a gas of $^{40}$K atoms trapped in this geometry were studied
\cite{Vogt2012}. The interaction strength between the two hyperfine
states and the temperature were varied in order to identify the
transition from the collisionless to the hydrodynamic regime in the
case of the quadrupole mode, and to confirm, in the case of the breathing 
mode, the dynamical scaling predicted a few years ago
\cite{Pitaevskii}. In a hydrodynamic picture, the damping of the
quadrupole mode is related to the shear viscosity of the 2D gas: in
particular, from the experimental results one can extract the
temperature dependence of the shear viscosity.

A number of theoretical studies dealing with this experiment have
already appeared
\cite{Bruun2012,Schafer2012,Enss2012,WuZhang2012_2d,Baur2013}. In
\Refs{Bruun2012,Schafer2012} the shear viscosity and spin-diffusion
coefficients were computed from kinetic theory in the hydrodynamic
regime. Surprisingly, it was found that the quantitative agreement
with data was better in the collisionless regime. In \Ref{Enss2012},
in-medium modifications of the scattering cross-section were included
in the calculation of the shear viscosity, leading to a maximum
damping rate as high as the experimental one under the assumption that
the hydrodynamic approach was valid. In \Ref{WuZhang2012_2d}, the
Boltzmann equation was numerically solved in the local relaxation-time
approximation (the relaxation time being calculated with the
free-space cross section). The authors found a reasonable agreement
with the experimental data in the case of moderate quantum degeneracy
and not too strong interactions, if the computed damping was shifted
upwards by a small constant value, introduced to account for
additional effects like trap anharmonicity that had to be there since
the experiment observed a finite damping of the dipole mode. In the
most recent work by Baur et al. \cite{Baur2013}, the in-medium
scattering cross-section was included into the Boltzmann equation
(as it was done before in 3D \cite{Riedl2008,Chiacchiera2009})
which was then solved in an approximate way by using the method of
second-order phase-space moments. They conclude that they can well 
describe the experimental results, apart from an offset in the damping 
rate (see caption of Fig. 2 of \Ref{Baur2013}).

In the case of collective modes in 3D Fermi gases, it was found that
in order to quantitatively reproduce the result of a numerical
solution of the Boltzmann equation \cite{Lepers2010}, the method of
second-order moments was insufficient and higher-order moments had to
be included. The inclusion of fourth-order moments improved a lot the
agreement between theory and experiment in the case of the quadrupole 
mode \cite{Chiacchiera2011}.
Furthermore, by using moments
up to third order, it was possible to describe the effects of the trap
anharmonicity (frequency shift and damping) on the sloshing mode
\cite{Pantel2012} in 3D. Higher-order moments were also used in the
context of 2D systems to describe collective modes in dipolar Fermi
gases \cite{Babadi2012}.

In the present work, we will extend the analysis of the paper by Baur
et al. \cite{Baur2013}. We will include all phase-space moments up to
fourth order and study the effect of a realistic form of the trap
potential having a gaussian shape with additional linear terms due to
gravity and magnetic-field gradients. All this causes some damping of
the quadrupole mode even in the non-interaction regime, but not as
strong as the one observed in the experiment. We will also discuss
other possible sources of damping like dephasing between different
slices and the time-of-flight (TOF) before the measurement of the
quadrupole moment, but they are all too weak to explain the data.

Our paper is organized as follows. In \Sec{sec:forma}, we briefly
summarize the formalism. In \Sec{sec:compWuZhang} we investigate the
convergence of the moments method to existing numerical solutions of
the Boltzmann equation. In \Sec{sec:compVogt}, we try to model as
closely as possible the experiment \cite{Vogt2012}, and in
\Sec{sec:conclusion} we draw our conclusions.

Throughout the paper, we use units with $\hbar=k_B=1$ ($\hbar=h/2\pi$
and $k_B$ being the reduced Planck constant and the Boltzmann
constant, respectively). In a harmonic trap with average frequency
$\wbarrad$, it is convenient to work with so-called ``trap units'',
e.g., the energy unit $\hbar\wbarrad$, the length unit $l_\ho =
\sqrt{\hbar/m\wbarrad}$, etc, $m$ being the atom mass. Furthermore,
the Fermi energy $E_F = k_BT_F$ and Fermi momentum $k_F$ of a 2D
trapped gas are defined by $E_F = \hbar\wbarrad\sqrt{N}$ and $k_F =
\sqrt{2mE_F}/\hbar$.
\section{\label{sec:forma}Formalism}
\subsection{Boltzmann equation in 2D}
We consider a two-component ($\uparrow, \downarrow$) Fermi gas of $N$
atoms of mass $m$ that can move only in two dimensions ($x, y$). For a
balanced mixture and ``spin''-independent modes, it is enough to
consider a single phase-space distribution function $f = f_\uparrow =
f_\downarrow$, normalized to $\int \dgammatext f(\rv,\pv)= N/2$, where
$\rv=(x,y)$, $\pv=(p_x,p_y)$. Averages are computed as
\begin{equation}
\ave{q}(t) = \frac{2}{N} \int \dgamma
f(\rv,\pv,t) q(\rv,\pv)\,.
\label{eq:average}
\end{equation}
At equilibrium, the distribution is given by the Fermi function
\begin{equation}
  f_\eq(\rv,\pv) = 
    \frac{1}{e^{\beta\left(\frac{\pv^2}{2 m}+\Vtrap(\rv)-\mu_0\right)} + 1},
\end{equation}
where $\beta=1/T$, $\Vtrap$, and $\mu_0$ are the inverse temperature,
the trap potential, and the chemical potential, respectively. Note
that we neglect a possible mean-field potential, since in
\Refs{Chiacchiera2009} and \cite{Pantel2012} we have shown (in the 3D
case) that it does not substantially affect the properties of the
collective modes.

For small deviations from equilibrium, the change of the phase-space 
distribution $\delta f=f-f_\eq$ can be written as \cite{Landau10}
\begin{equation}
\delta f(\rv,\pv,t) = f_\eq \fbar_\eq\Phi(\rv,\pv,t)\,,
\end{equation}
where $\fbar = 1-f$. The prefactor $f_\eq(1-f_\eq)$ takes care of the
rapid variation of $\delta f$ around the Fermi surface, so that $\Phi$
can be considered a smooth function of $\rv$ and $\pv$. Then the
linearized Boltzmann equation becomes
\begin{multline}
  f_\eq\fbar_\eq \Big(\dot{\Phi}
  +\Big\{\Phi,\frac{p^2}{2m}+\Vtrap(\rv)\Big\}
  + \beta \frac{\pv}{m} \cdot \nablav_r \delta V \Big)
  = -I[\Phi]\,,
  \label{eq:BoltzLin}
\end{multline}
where $\{\cdot,\cdot\}$ denotes the Poisson bracket and $\delta V$ is
a perturbation of the trap potential used to excite the collective
mode. Usually we consider a perturbation of the form of a $\delta$
pulse,
\begin{equation}
\delta V(\rv,t)=\delta(t) \hat{V}(\rv)\,.
\label{eq:Vpulse}
\end{equation}

The linearized collision integral $I[\Phi]$ reads
\begin{multline}
  I[\Phi] = \int \frac{d^2p_1}{(2\pi)^2} \int_0^{2\pi} d \theta\,
  \frac{d\sigma^{2D}}{d\theta} \frac{|\pv-\pv_1|}{m} f_\eq f_{\eq 1}
  \fbar_\eq^\prime \fbar_{\eq 1}^\prime\\
  \times (\Phi+\Phi_1-\Phi^\prime-\Phi_1^\prime)\,,
\label{eq:Collisionterm}
\end{multline}
where the short-hand notation $f = f(\rv,\pv)$, $f_1 = f(\rv,\pv_1)$,
$f^\prime = f(\rv,\pv^\prime)$, etc., has been used. Momentum and
energy conservation imply $\pv+\pv_1 = \pv^\prime+\pv_1^\prime$ and
$|\pv-\pv_1| = |\pv^\prime-\pv_1^\prime|$, and $\theta$ denotes the
scattering angle between $\pv-\pv_1$ and
$\pv^\prime-\pv_1^\prime$.
\subsection{Cross section}
In 2D, the differential cross section that enters
\Eq{eq:Collisionterm} has the dimension of a length. In free space, it
is given by \cite{Adhikari1986}
\begin{equation}
\frac{d \sigma_0^{2D}}{d \theta}=
  \frac{2\pi}{q}\frac{1}{\ln^2{(q^2a_{2D}^2)}+\pi^2}
\end{equation}
where $\vek{q}=(\pv-\pv_1)/2$ is the momentum of the atoms in the
center-of-mass frame, $\theta$ is the scattering angle, and $a_{2D}$
is the 2D scattering length \cite{Petrov2001} which is related to the 
dimer binding energy $E_B$ by $|E_B| = 1/(ma_{2D}^2)$. At finite density,
 the scattering cross section is obtained from the in-medium $\calT$ matrix
as
\begin{equation}
\frac{d \sigma^{2D}}{d \theta}= \frac{m^2}{8\pi q} |{\mathcal
  T}(k,\omega)|^2\,,
\end{equation}
where $\vek{k}=\pv_1+\pv_2$ is the total momentum of the pair and
$\omega = k^2/(4m)+q^2/m-2\mu_\mathit{loc}(\rv)$ is the total energy
of the pair. Here we use the in-medium $\calT$ matrix from
\Ref{Enss2012}, which is calculated in the non self-consistent ladder
approximation,
\begin{equation}
  \mathcal T^{-1}(k,\omega) = \frac{m}{4\pi} \ln 
  \frac{-1/(ma_{2D}^2)} {\omega+2\mu_{\loc}(\rv)-\frac{k^2}{4m}+i0}
  + I_\mathit{med}.
\end{equation}
The in-medium term $I_\mathit{med}$ incorporates Pauli blocking in the
intermediate states of the ladder and is given by
\begin{align*}
  I_\mathit{med} & = \int \frac{d^2p}{(2\pi)^2} \,
  \frac{f_{\eq}(\rv,\pv)+f_{\eq}(\rv,\kv-\pv)}
  {\omega+2\mu_{\loc}(\rv)-\frac{p^2}{2m}-\frac{(\kv-\pv)^2}{2m}+i0} \\
  & = \int \frac{dp\,p}{2\pi}\, \frac{f_{\eq}(\rv,\pv) \sgn(\Omega)}
  {\sqrt{(\Omega+i0)^2-k^2p^2/(4m^2)}}
\end{align*}
with $\Omega = \omega/2+\mu_{\loc}(\rv)-k^2/(4m)-p^2/(2m)$. In the
second line the angular integration has been performed, while the
radial integral is known analytically only at zero temperature. The
in-medium $\calT$ matrix depends on $k$, $\omega$, $a_{2D}$, $T$, and
the local chemical potential $\mu_{\loc}(\rv) = \mu_0 - \Vtrap(\rv)$.

\begin{figure}
  \includegraphics[width=7cm]{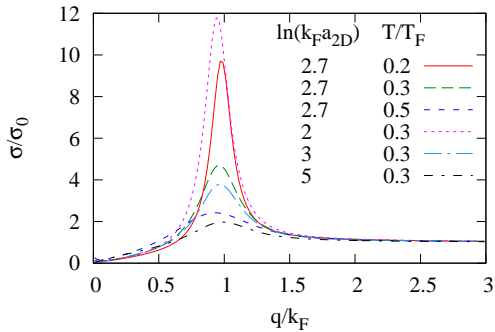}
  \caption{(Color online) Ratio of in-medium and free-space cross-section for total
    momentum $k$=0 as a function of the relative momentum $q$ for
    various combinations of interaction strength 
    $\ln(k_Fa_{2D}) = \ln(2E_F/|E_B|)/2$ and
    temperature $T/T_F$.}
  \label{fig:xsec}
\end{figure}

In \Fig{fig:xsec} 
we show the ratio of the in-medium cross section and the free-space
one as a function of the relative momentum for a pair with $k=0$ for
different values of interaction strength and temperature. At fixed
interaction strength, there is a strong enhancement of the cross
section when the temperature $T/T_F$ decreases, a precursor effect of
superfluidity (cf. Fig.\@ 2 of \Ref{Chiacchiera2009} for the 3D case,
see also \Ref{AlmRoepke} for the analogous effect in nuclear
matter). This enhancement is most pronounced in the strongly
interacting regime [small $\ln(k_Fa_{2D})$]. Note that,
  throughout this paper, we consider only the fermionic regime in the
  normal phase, i.e., the case $\ln(k_Fa_{2D}) > 0$ at temperatures
  above the superfluid transition temperature.
\subsection{Method of phase-space moments}
\label{sec:moments}
We look for approximate solutions of the Boltzmann equation using the
method of phase-space moments. By fixing the functional form of
$\Phi$ as
\begin{equation}
\Phi(\rv,\pv,t)=\sum_{j=1}^n c_j(t)\phi_j(\rv,\pv)\,,
\end{equation}
the basis functions $\phi_j(\rv,\pv)$ being monomials in $\rv$ and
$\pv$, one can obtain a closed set of $n$ coupled equations for the
coefficients $c_j$ by multiplying \Eq{eq:BoltzLin} by $\phi_i$ and
integrating over phase space. After Fourier transformation, the
equations become algebraic and can be written in matrix form as
$\sum_{j=1}^n A_{ij}(\omega) c_j(\omega)= a_i$, where $A$ is related
to the transport and collision part and $a$ to the perturbation
$\delta V$, see \Refs{Chiacchiera2011,Pantel2012}. Once the
coefficients $c_i$ are found, the deviation of any one-body observable
from its equilibrium value, $\delta\ave{q} = \ave{q}-\ave{q}_\eq$, can
be expressed in the form $\delta\ave{q} = \sum_{i=1}^n b_ic_i$, 
$b_i$ being appropriate projections of the observable 
on the basis.

The choice of the $\phi_i$, of the excitation $\hat{V}$ and of the
observable $\ave{q}$ depends on the mode one is interested in. In
general, the response function $\ave{q}_\pulse(\omega) =
\delta\ave{q}(\omega)$ for the $\delta$-pulse excitation,
\Eq{eq:Vpulse}, has $n$ poles at complex frequencies $\omega_i$. In
simple cases, the real and imaginary parts of $\omega_i$ can directly
be interpreted as the frequency and damping rate of the collective
mode, as it was done, e.g., in \Ref{Baur2013}. In general, however, it
is necessary to analyse the full response function in order to extract
the frequency $\omega$ and damping rate $\Gamma$ of the collective
mode \cite{Chiacchiera2011}.

In the present paper, we will follow as closely as possible the
experimental procedure. Note that in real experiments the mode
is not excited by a $\delta$ pulse of the form \Eq{eq:Vpulse}, but the
perturbation is adiabatically switched on and then suddenly switched
off at $t=0$. The corresponding response $\ave{q}_\step$ can easily be
calculated (see appendix for more details) if the response $\ave{q}_\pulse$
for the $\delta$ pulse is known. Then we fit the response $\ave{q}_\step(t)$
with a function of the form
\begin{equation}
Q(t) = A e^{-\Gamma t} \cos(\omega t+\varphi)+B e^{-\gamma t}\,,
\label{eq:Qfit}
\end{equation}
as it is done in the analysis of the experimental data, in order to
determine $\omega$ and $\Gamma$.
\section{Comparison between moments method and numerical calculations}
\label{sec:compWuZhang}
In this section we will discuss the quadrupole mode in an isotropic
harmonic trap, $\Vtrap = \frac{1}{2}m\wrad^2r^2$. The minimal ansatz
function $\Phi$ is in this case given by
\begin{equation}
  \Phi = c_1(x^2-y^2) + c_2(p_x^2-p_y^2)+c_3(xp_x-yp_y)\,.
\end{equation}
The excitation operator is $\hat{V} \propto x^2-y^2$ and the
observable is the quadrupole moment of the cloud, $\ave{q} =
\ave{x^2-y^2}$. This defines the method of moments at second order.
Within this approximation, the frequency and damping rate of the
quadrupole mode depend only on a single parameter, the average
relaxation time $\tau$ \cite{Baur2013}. In the hydrodynamic limit
$\tau \to 0$, one finds $\omega \to \sqrt{2}\wrad$ and $\Gamma\to
0$. In the collisionless limit $\tau\to\infty$, one finds $\omega \to
2\wrad$ and $\Gamma\to 0$. The maximum damping of $\Gamma\sim
0.354\wrad$ is reached for $\tau\sim 0.471/\wrad$. Hence, whether one
includes a medium-modified cross section or not changes only the
dependence of $\tau$ on $a_{2D}$, $T$ etc., but it cannot lead to a
damping rate higher than $0.354\wrad$, which is far below the observed
maximum damping of $\sim 0.6\wrad$ \cite{Vogt2012}.

In the 3D case, we have shown by comparing with numerical simulations
that the second-order method overestimates the collision effects
\cite{Lepers2010}. In the 2D case, the results of numerical
calculations are already available \cite{WuZhang2012_2d}. Although
they still use a relaxation-time approximation, they include the
essential effect that is missing in the second-order method, namely
the position-dependence of the local relaxation time
$\tau(\rv)$. Since $\tau(\rv)$ depends on collisions, it strongly
increases if one goes from the trap center (high density) to the
surface of the gas (low density).

As discussed in \Refs{Lepers2010} and \cite{Chiacchiera2011}, this
effect is automatically taken into account if one extends the method
of moments to higher orders. As in the 3D case, we will include all
the relevant moments up to fourth order
\footnote{The term
  $\rv\cdot\pv(xp_x-yp_y)$ that was present in Eq.\, (D1) of
  \Ref{Lepers2010} is not needed in 2D since it is equal to
  $\frac{1}{2}[r^2(p_x^2-p_y^2)+p^2(x^2-y^2)]$.}, i.e.
\begin{multline}
\Phi=
c_1(x^2-y^2)+c_2(p_x^2-p_y^2)+c_3(x p_x-y p_y)\\ 
+c_4r^2(x^2-y^2)+c_5r^2(p_x^2-p_y^2)+c_6r^2(xp_x-yp_y)\\
+c_7p^2(x^2-y^2)+c_8p^2 (p_x^2-p_y^2)+c_9p^2(x p_x-y p_y)\\
+c_{10}\rv\cdot\pv(x^2-y^2)+c_{11}\rv\cdot\pv(p_x^2-p_y^2)\,.
\end{multline}

\begin{figure}
  \includegraphics[width=6cm]{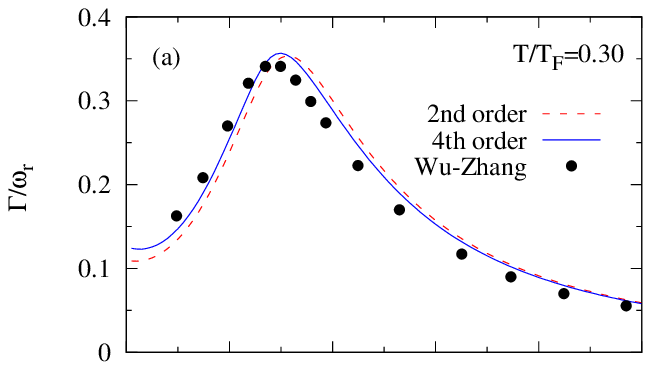}\\[-2mm]
  \includegraphics[width=6cm]{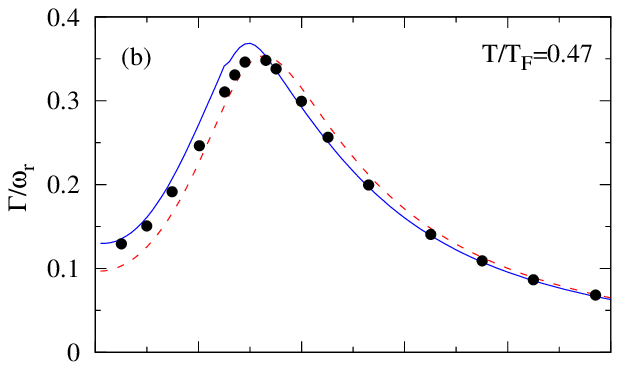}\\[-2mm]
  \includegraphics[width=6cm]{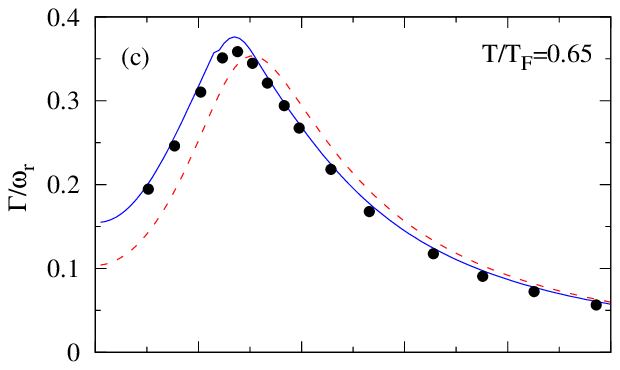}\\[-2mm]
  \includegraphics[width=6cm]{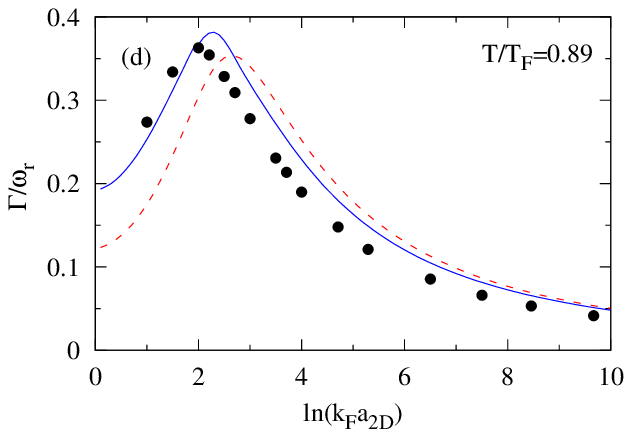}
  \caption{\label{fig:GammaWuZhang}(Color online) Damping rate $\Gamma$ of the
    quadrupole mode as a function of the interaction strength for a
    system of $N=3500$ $^{40}$K atoms with the free-space cross
    section in a harmonic isotropic trap with $\wrad = 2\pi\times 125$
    Hz at different temperatures: (a) $T/T_F = 0.3$, (b) 0.47, (c)
    0.65, (d) 0.89. Dashed lines: second-order moments, solid lines:
    fourth order moments, points: numerical results by Wu and Zhang
    \cite{WuZhang2012_2d}.}  \includegraphics[width=6cm]{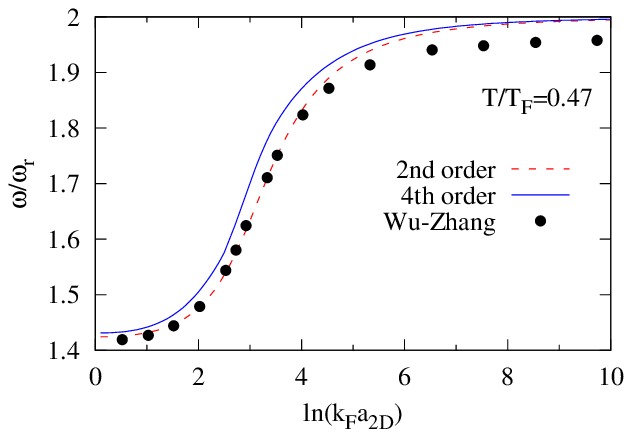}
  \caption{\label{fig:wWuZhang}(Color online) Frequency $\omega$ of 
    the quadrupole
    mode as a function of the interaction strength. The parameters of
    the system and the meaning of the different lines are the same as
    in \Fig{fig:GammaWuZhang}(b).}
\end{figure}

Let us now compare the second- and fourth-order results with the numerical
results of \Ref{WuZhang2012_2d}. The damping rate $\Gamma$ as function
of the interaction strength $\ln(k_F a_{2D})$ is shown for different
temperatures in \Fig{fig:GammaWuZhang}.
For the sake of comparison, we used the free-space cross section in our
calculation, and we also removed the constant shift of $0.05\wrad$
from the numerical damping rates that was added in
\Ref{WuZhang2012_2d} to account in a simple way for the anharmonicity
of the experimental trap potential (anharmonicity effects will be
discussed in detail in the next section). We observe that within the
second-order moments method (dashed lines) the transition from
hydrodynamic to collisionless behavior, i.e., the maximum of $\Gamma$,
lies at slightly weaker interaction [larger $\ln(k_Fa_{2D})$] than
within the numerical calculation (points). This is in line with our
results for 3D, where the second-order method overestimates the
collision effects, too \cite{Lepers2010}. The fourth-order results are
in very good agreement with the numerical ones, especially at higher
temperature [\Figs{fig:GammaWuZhang}(c) and (d)], where the difference
between the second- and fourth-order results becomes more pronounced.

In \Fig{fig:wWuZhang} we also show the frequency $\omega$ as function
of $\ln(k_Fa_{2D})$ for the temperature $0.47T_F$ for which numerical
results are available. At first glance it looks as if the numerical
result was in better agreement with the second-order calculation than
with the fourth-order one. However, one sees that in the weakly
interacting regime the numerical frequency stays systematically $2\%$
below both the second- and the fourth-order results. If we multiply
the numerical frequencies by 1.02, they lie between the second- and
fourth-order results in the range $\ln(k_Fa_{2D})\ge 2$.

In conclusion, the second-order method overestimates the role of
collisions. Especially at higher temperatures, the inclusion of
fourth-order moments reduces the effects of collisions and
significantly improves the agreement between the damping rates
obtained within the method of moments and those obtained from a
numerical calculation. However, at low temperatures, the corrections
due to fourth-order moments are small.
\section{Comparison with experiment}
\label{sec:compVogt}
\subsection{Effect of the in-medium cross section}
From now on, we will concentrate on results obtained with the
fourth-order method, and compare them with the experiment of
\Refs{Vogt2012,Baur2013}
\footnote{In fact, the data presented in \Refs{Vogt2012,Baur2013} 
result from different analyses of the same experiment. 
In the more recent \Ref{Baur2013} the analysis has been refined 
for $T/T_F=0.47$.}. 
As a first step, we approximate the
experimental system again by an isotropic harmonic trap with $\wrad =
2\pi\times 125$ Hz. In contrast to the preceding section, we will now
include the in-medium cross section into the collision term. Within
the moments method this is feasible, while it would be tremendously
time-consuming in a numerical simulation like that of
\Ref{WuZhang2012_2d}.

\begin{figure}
  \includegraphics[width=6cm]{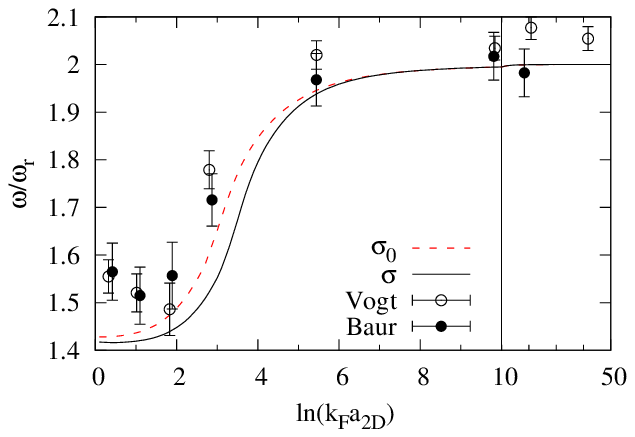}
  \includegraphics[width=6cm]{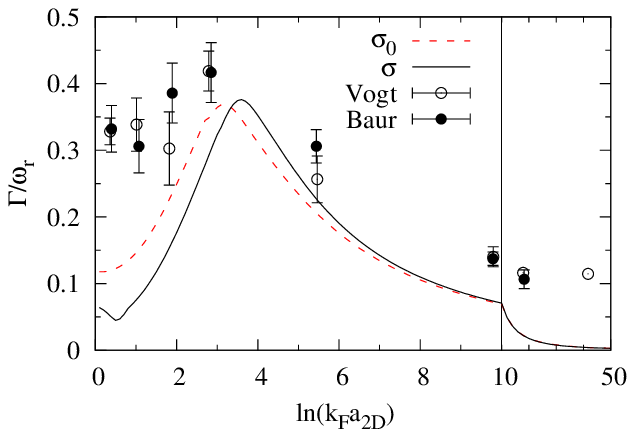}
  \caption{(Color online) Frequency (top) and damping rate (bottom) 
    of the quadrupole mode in a harmonic isotropic 
    trap containing $N=4300$ atoms at $T/T_F = 0.47$. 
    The dashed and solid lines represent fourth-order results
    obtained with the free-space and in-medium cross sections,
    respectively. The experimental data are taken from \Refs{Vogt2012}
    and \cite{Baur2013}.}
  \label{fig:Vogtharmonic}
\end{figure}

In \Fig{fig:Vogtharmonic}, we show the frequency and damping rate of
the quadrupole mode as functions of the interaction strength for the
case of $N=4300$ atoms ($E_F = h\times 8.2$ kHz) at $T/T_F = 0.47$.
Since the in-medium cross section is enhanced, its main effect is that
the system is more hydrodynamic (weaker damping) for strong
interactions [small values of $\ln(k_Fa_{2D})$] and the transition to
the collisionless regime (maximum damping) takes place at weaker
interactions [higher values of $\ln(k_Fa_{2D})$] than with the free-space
cross section. A similar effect of the in-medium cross section was
already found in \Ref{Baur2013} within the second-order method
(cf. Fig.\, 2 of \Ref{Baur2013}). Concerning the agreement with the
experimental data, one notes that the frequencies are qualitatively
correctly described, but the rise from the hydrodynamic to the
collisionless frequency happens at too weak interaction, and the
disagreement gets worse if the in-medium cross section is used instead
of the free-space one. The theoretical results for the damping are
significantly too weak in almost the whole range of
$\ln(k_Fa_{2D})$, especially in the very weakly interacting
  regime [$\ln(k_Fa_{2D})\gtrsim 10$].
\subsection{Realistic trap potential}
As mentioned before, the damping of the quadrupole mode one obtains
with the second-order method cannot exceed $0.354 \wrad$,
independently of the cross section that is used in the collision
term. In the fourth-order method, the damping can be somewhat
stronger, but it stays far below the maximum damping $0.6\wrad$ that
was observed in one case in the experiment by Vogt et al.
\cite{Vogt2012}. Furthermore, in this experiment, the quadrupole mode
remains damped in the limit of vanishing interaction strength. This
clearly shows that there are other sources of damping than the
collision term, for instance the anharmonicity of the trap potential
and the broken rotational symmetry. (By the way, damping of collective 
modes in the non-interacting gas was also observed in 3D, 
for instance in \Ref{Kinast2004}
where it was consistent with the trap anharmonicity.) 

In \Ref{Pantel2012}, we showed that in 3D the damping of the sloshing
mode in an anharmonic potential could be described within the method
of moments once moments of higher order were included. Therefore we
expect that the inclusion of higher-order moments in the description
of the quadrupole mode will also allow us to describe its additional
damping in an anharmonic trap.

The breaking of rotational invariance leads to a coupling of modes of
different multipolarity, that can also result in an additional
damping. For instance, in \Ref{Baur2013}, the coupling of quadrupole
and monopole (breathing) modes caused by the small ellipticity of the
trap potential was studied. In an elliptic trap, the monopole mode and
the two degenerate quadrupole modes (in 2D) are replaced by three new
eigenmodes that have all different frequencies. Notice that a beat
caused by the superposition of two eigenmodes with slightly different
frequencies looks like a damping if the oscillation is only observed
during a short time interval, as it is usually the case.

In the experiment \cite{Vogt2012} there is another effect that might
play a role. Since the $z$ direction of the laser beam generating the
potential is horizontal, the additional gravitational potential shifts
the minimum of the potential downwards. While this would not have any
effect in a purely harmonic potential, it leads in the anharmonic case
to a potential that is no longer symmetric about its minimum. As a
consequence, modes with opposite parity (e.g., sloshing and breathing)
will be coupled. Actually, the symmetry in $x$ direction is broken,
too, because of the presence of magnetic field gradients that shift
the minimum in both $x$ and $y$ directions \cite{Vogt_Thesis}.

We write our model potential as
\begin{equation}
\tilde{V}_\trap(\tilde{\rv}) = -V_0 e^{-2(\tilde{x}^2/w_x^2+\tilde{y}^2/w_y^2)} -
  (m\vek{g}+\mu \nablav B) \cdot\tilde{\rv}\,,
\label{eq:Vexp}
\end{equation}
where $V_0$ is the depth of the Gaussian potential, $w_x$ and $w_y$
are the waists of the laser beam in $x$ and $y$ directions, $\vek{g} =
-g\vek{e}_y$ is the gravitational acceleration ($g=9.81$ m$/$s$^2$),
$\mu$ is the magnetic moment (approximately equal to the Bohr magneton
$\mu_B$ in the case of alkali atoms), and $B=|\vek{B}|$ is the
strength of the magnetic field. For the sake of simplicity, we shift
the minimum of the potential to the origin by defining $\rv =
\tilde{\rv}-\rv_0$ and
\begin{equation}
\Vtrap(\rv)=\tilde{V}_\trap(\rv_0+\rv)-\tilde{V}_\trap(\rv_0)\,,
\end{equation}
where $\rv_0 = (x_0,y_0)$ is related to $\vek{g}$ and $\nablav B$ by
\begin{equation}
m g_i+\mu\nabla_i B = \frac{4V_0r_{0i}}{w_i^2}
e^{-2(x_0^2/w_x^2+y_0^2/w_y^2)} \quad(i = x,y)\,.
\end{equation}
The average trap frequency $\wbarrad$ can be obtained from
\begin{multline}
m\wbarrad^2 = \begin{vmatrix}
  \frac{\partial^2 V_T}{\partial x^2}&\frac{\partial^2 V_T}{\partial x\partial y}\\
  \frac{\partial^2 V_T}{\partial x\partial y}&\frac{\partial^2 V_T}{\partial y^2}
\end{vmatrix}_{\rv=0}^{1/2}\\
  = \frac{4 V_0}{w_xw_y}\sqrt{1-\frac{4x_0^2}{w_x^2}-\frac{4y_0^2}{w_y^2}}
     e^{-2(x_0^2/w_x^2+y_0^2/w_y^2)}\,.
\label{eq:averageomegar}
\end{multline}

\begin{figure}
  \includegraphics[width=7cm]{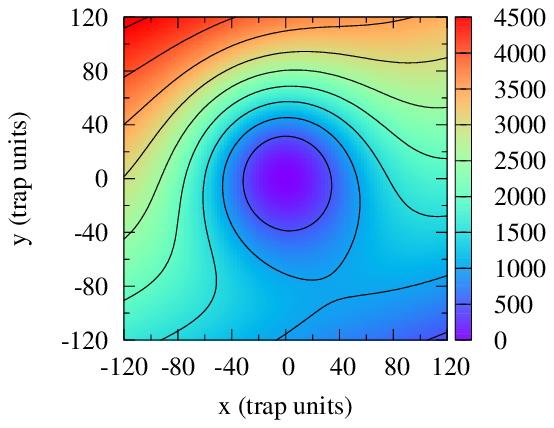}\\
  \includegraphics[width=7cm]{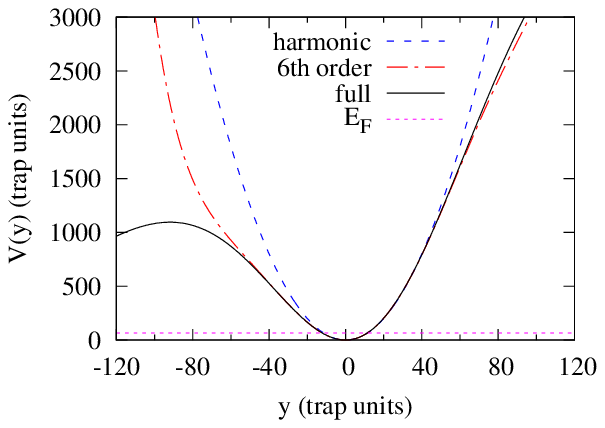}
  \caption{(Color online) Trap potential in units of $\hbar\wbarrad$ 
    as function of $x$
    and $y$ in units of $l_\ho$ (top) and as function of $y$ for $x=0$
    (bottom). We have used $\wbarrad = 2\pi\times 125$ Hz \cite{Vogt2012}
    and $w_x=139$ $\mu$m, $w_y=142$ $\mu$m, $\nabla_x B = 3.2$ G$/$cm
    and $\nabla_y B = -0.75$ G$/$cm \cite{Vogt_Thesis} (as a
    consequence, $V_0 = 2617\hbar\wrad$, $x_0 = 4.8 l_\ho$, and $y_0
    = -12.2 l_\ho$). Solid line: potential according to \Eq{eq:Vexp},
    dashed line: harmonic approximation, dash-dotted line: Taylor
    expansion up to sixth order. The Fermi energy $E_F$ corresponding
    to $N = 4300$ atoms is shown as the dotted line.}
  \label{fig:vtrap}
\end{figure}

Using the parameters of the experiments \cite{Vogt2012,Vogt_Thesis},
we obtain the potential shown in the upper panel of
\Fig{fig:vtrap}.
As one can clearly see, the principal axes of the potential near the
minimum are not aligned with the $x$ and $y$ axes. The trap
frequencies along the principal axes are split by approximately 5\%.

As explained in \Ref{Pantel2012}, it is strictly speaking not possible
to calculate the chemical potential $\mu_0$ as a function of the
particle number $N$ if the potential does not go to $+\infty$ for
$r\to\infty$. We avoid this problem in the same way as in
\Ref{Pantel2012} by expanding the potential up to sixth order (i.e.,
keeping terms $\propto x^ky^l$ with $k+l\le 6$) around $r=0$. In the
present case, this expansion is very accurate up to energies of about
ten times the Fermi energy. This is illustrated in the lower panel of
\Fig{fig:vtrap}.

As mentioned above, the asymmetry of the potential leads to a coupling
between all kinds of modes, even those of different parity. If we want
to describe this in the framework of the moments method, we have to
make the most general ansatz, i.e., include all possible moments up to
a given order. Our ansatz for $\Phi$ contains now 70 terms (1 of
zeroth order, 4 of first order, 10 of second order, 20 of third order
and 35 of fourth order) and reads
\begin{equation}
\Phi(\rv,\pv,t) = \sum_{k+l+m+n\le 4} c_{klmn}(t)\phi_{klmn}(\rv,\pv) 
\end{equation}
where $k,l,m$ and $n$ are non-negative integers and
\begin{equation}
\phi_{klmn}(\rv,\pv) = x^k y^l p_x^m p_y^n\,.
\end{equation}
The zeroth-order (constant) term is necessary for the conservation of
the particle number during the oscillation \cite{Babadi2012}.
This choice of $\Phi$, together with the same excitation 
operator and observable as before, $\hat{V} \propto x^2-y^2 = q$, 
define our method at fourth order.

\begin{figure}
  \includegraphics[width=6cm]{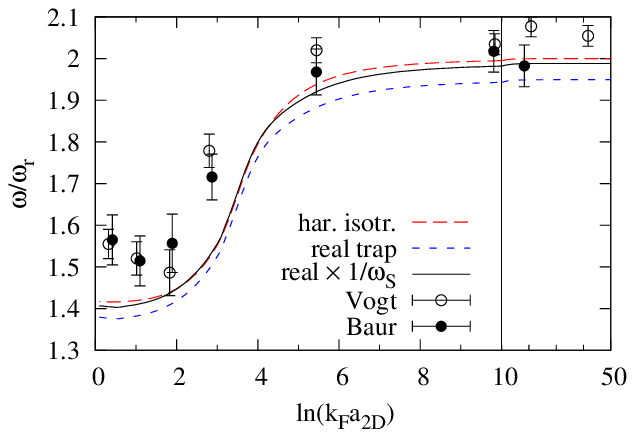}
  \includegraphics[width=6cm]{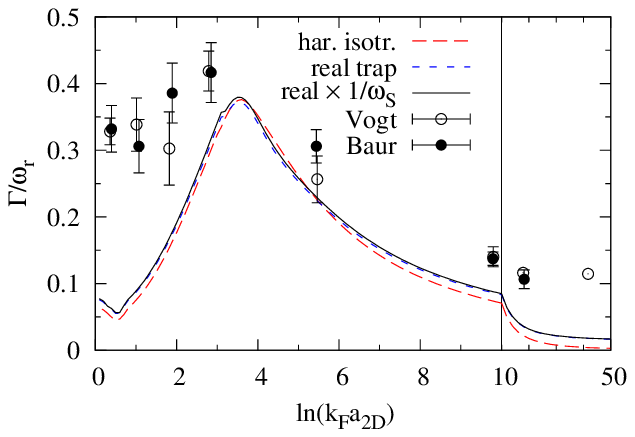}
  \caption{(Color online) Fourth-order results with in-medium cross
    section for the
    frequency (top) and damping rate (bottom) of the quadrupole mode
    in a trap containing $N=4300$ atoms at $T/T_F = 0.47$. Long and
    short dashes represent, respectively, results for a harmonic
    isotropic trap and for the realistic trap shown in
    \Fig{fig:vtrap}. The solid lines are the results for the realistic
    trap, normalized, as in the experiment, by the average sloshing
    frequency $\omega_S \approx 0.98\omega_\rad$ instead of $\omega_\rad$
    given by \Eq{eq:averageomegar}. The experimental data are taken
    from \Ref{Vogt2012} (empty circles) and \cite{Baur2013} (filled
    circles).}
  \label{fig:Vogtanharmonic}
\end{figure}

In \Fig{fig:Vogtanharmonic}
we display results obtained within the full calculation (moments up to
fourth order, in-medium cross section) for the case of a harmonic
isotropic trap (long dashes) and for the realistic trap (short dashes
and solid lines). One can see that the damping in the weakly
interacting limit [$\ln(k_Fa_{2D}) \gtrsim 10$] is significantly
enhanced in the realistic trap. Actually, the main reason for the
additional damping $\Gamma$ is not the anharmonicity but the
ellipticity of the trap. As discussed in the beginning of this
subsection, this effect was already considered in \cite{Baur2013} but
not analyzed in the same way. In our analysis, the beat caused by the
two quadrupole modes that do no longer have the same frequencies
results in a finite damping rate $\Gamma$ when the response is fitted
with a single damped cosine function, \Eq{eq:Qfit}, on a relatively
short time interval. However, the effect is far too weak to explain
the experimentally observed damping. At smaller values of
$\ln(k_Fa_{2D})$, the damping is not substantially modified by the
anharmonicity and ellipticity of the trap.

The main effect of the anharmonicity is to reduce the frequency of the
quadrupole mode (cf.\ the short and long dashed lines in the upper
panel of \Fig{fig:Vogtanharmonic}). This looks incompatible with the
data. However, if we normalize our quadrupole mode frequency, as in
the experiment, by the average sloshing frequency $\omega_S =
0.98\omega_\rad$ instead of the average trap frequency defined by the
second derivatives at the minimum, \Eq{eq:averageomegar}, this effect
disappears (solid line) because the anharmonicity reduces quadrupole
and sloshing frequencies by approximately the same factor.

\begin{figure}
  \includegraphics[width=6cm]{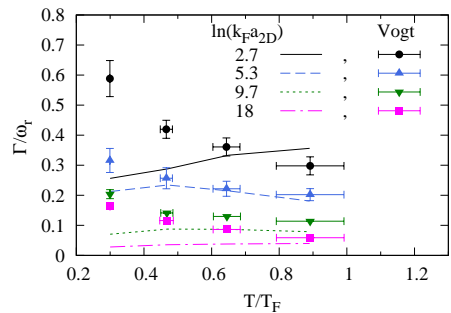}
  \caption{(Color online) Damping rates of the quadrupole mode in 
    a realistic trap as
    functions of temperature for different interaction strengths. The
    lines represent fourth-order results with in-medium cross section
    in the realistic trap potential. The data points are taken from
    \Ref{Vogt2012}. The particle numbers for $T = 0.3, 0.47, 0.65$,
    and $0.89\, T_F$ are, respectively, $N = 2620, 4300, 5180$, and
    $5300$, corresponding to the Fermi energies given in
    \Ref{Vogt2012}.}
  \label{fig:3aVogt}
\end{figure}

The damping rates of the quadrupole mode for other temperatures are
shown in \Fig{fig:3aVogt}.
To be consistent with the experiment, for each temperature the
calculations are performed with a different value of $N$ (see caption
of \Fig{fig:3aVogt}). The agreement between theory and data varies
from each data point to the other, but two clear trends are visible:
First, the experimentally observed damping is much stronger than the
theoretical result at low temperature, $T/T_F = 0.3$, for all values
of the interaction strength. Second, the experimental damping in the
weakly interacting case, $\ln(k_Fa_{2D}) \ge 9.7$, is also stronger
than the theoretical one for all temperatures. Surprisingly, the
experimental damping rate in the weakly interacting limit decreases
with increasing temperature, while one would expect the opposite
behavior if this damping was related to anharmonicity effects [cf.
dashed-dotted line corresponding to $\ln(k_Fa_{2D})=18$].
\subsection{Other possible effects}
As we have seen, the agreement between theory and data is not
satisfactory. While one maybe cannot trust the Boltzmann equation in
the limit of strong interaction, it should at least be valid at large
$\ln(k_Fa_{2D})$, but even there a systematic disagreement between the
theoretical and experimental damping rates persists. Possible effects
one might think of are:
\begin{enumerate}
\item[(a)] The excitation $\Vhat$ is not of the form $x^2-y^2$, but it
  consists in squeezing the laser in one direction and stretching it
  in the other direction. This leads to anharmonic terms. In addition,
  it shifts the minimum of the potential and thereby excites not only
  the quadrupole but also the sloshing mode.
\item[(b)] The observable $q$ is not $x^2-y^2$, but it is the
  quadrupole moment of the cloud after a free expansion during a time
  of flight (TOF) $t_{TOF} = 12$ ms. This can easily be modeled, one
  just has to replace $x$ by $x+t_\TOF p_x/m$ and analogously for $y$.
\item[(c)] In the experiment, there is not a single 2D gas, but about
  30 layers (``pancakes'') containing different particle numbers. The
  measured response is the sum of the responses of all of these
  layers. In \Ref{Vogt2012} it was suggested that the dephasing
  between the different layers might be an explanation of the observed
  damping.
\end{enumerate}

The effects (a) and (b) do not change the eigenvalues, i.e., the poles
$\omega_i$ of the response function in the complex plane, but they
change the relative weight of the different eigenvalues and therefore
have some effect if one determines the quadrupole frequency $\omega$
and damping rate $\Gamma$ by fitting the response function.  In this
respect, we note that, since with the full perturbation also a
sloshing mode is excited, the fitting function has to be extended to
take it into account. We studied in detail the case $T/T_F = 0.47$ and
found that in the collisionless regime [$\ln(k_Fa_{2D})\gtrsim 7$] the
results of the fits are not significantly changed. In the transition
region from the hydrodynamic to the collisionless regime around
$\ln(k_Fa_{2D}) = 4$ the effect (a) tends to increase $\Gamma$ (by
$\lesssim 10\%$) while (b) reduces it (by $\lesssim 10\%$), so that
the net effect is even smaller. In the strongly interacting
(near-hydrodynamic) regime [$\ln(k_Fa_{2D}) \lesssim 2$] the
Fermi-surface deformation gets so weak that the corresponding
quadrupole moment after the TOF is comparable with the quadrupole
moment of the cloud before the TOF. Since both oscillate out of phase,
the resulting amplitude can become very weak and the fit for the
determination of $\omega$ and $\Gamma$ fails. In all three cases, the
result of the fit depends very sensitively on details such as how the
center-of-mass motion is taken into account.

We also studied (c) the possible dephasing of the different layers.
When summing up the responses of layers having a distribution of
particle numbers as shown in Fig. 3.7(a) of \Ref{Froehlich_Thesis}, we
found that the total response is strongly dominated by the central
layers having the largest numbers of particles, since these have also
the largest radii. As a consequence, the effect of the peripheral
layers on the fitted frequency and damping rate is very weak. For
example, we studied the case $\ln(k_Fa_{2D}) = 3$, $T/T_F =
0.28$. Since this case is right in between the hydrodynamic and the
collisionless regimes, the frequency is supposed to depend strongly on
the parameter $\ln(k_Fa_{2D})$ that changes from one layer to the next
because of the different particle number in each layer. One might
therefore think that the dephasing could be important. Nevertheless we
found that by summing the responses of all the layers [weighting each
  response with the particle number of the corresponding layer to
  compensate the factor $1/N$ in \Eq{eq:average}] the frequency and
damping rate changes only by $\sim 2\%$.
\section{Conclusions}
\label{sec:conclusion}
We studied the quadrupole mode of a normal-fluid 2D trapped Fermi gas
in the framework of the Boltzmann equation. The Boltzmann equation was
solved approximatively within the method of phase-space moments. We
showed that by including moments of up to fourth order in $\rv$ and
$\pv$, we could nicely reproduce the results of the numerical study of
\Ref{WuZhang2012_2d}. In contrast to the 3D case \cite{Lepers2010},
the second-order moments alone were already in good agreement with the
numerical results, and the effect of the fourth-order moments was
quite small.

In order to compare with the experimental data of \Refs{Vogt2012} and
\cite{Baur2013}, we then included the in-medium cross section,
calculated within the ladder approximation \cite{Enss2012}, into the
collision integral. In \Ref{Enss2012}, a rough estimate based on the
shear viscosity of the uniform gas suggested that the inclusion of the
in-medium cross section instead of the free-space one could result in
a much stronger damping of the quadrupole mode. However, in agreement
with \cite{Baur2013}, we found that the effect of the in-medium cross
section was much less dramatic and consisted mainly in shifting the
transition from the hydrodynamic to the collisionless regime to
slightly weaker interactions or higher temperatures. The strong
damping rates observed in the experiment for $\ln(k_Fa_{2D}) = 2.7$ at
$T/T_F = 0.3$ and $0.47$ cannot be reproduced by our calculation.

There is also a strong discrepancy between theoretical and
experimental damping rates in the (almost) collisionless regime. In an
attempt to reconcile the theoretical results for the damping with the
much stronger damping observed in the experiment in this regime, we
included also the anharmonic shape of the experimental trap potential
into our calculation. In \Ref{Pantel2012}, we were able to explain in
this way the experimentally observed damping of the sloshing mode in
3D. In the present 2D case, however, it turned out that the
anharmonicity effects were very weak and did not substantially
increase the damping of the quadrupole mode. Other effects, such as
the expansion of the cloud or the summation over many 2D gases in the
optical lattice, were not able to explain the experimental data
either.

In the strongly interacting regime ($0 \lesssim \ln(k_Fa_{2D}) \lesssim 1$),
 it is maybe not so surprising that the Boltzmann equation does not 
reproduce the experimental data, since one might still be at the edge 
of the pseudogap phase \cite{Feld2011} where the quasiparticle 
picture breaks down. However, it is very puzzling that it also 
fails to describe the data in the weakly interacting case. Actually,
in the experiment \cite{Vogt2012}, a finite damping of the quadrupole
mode persists even in the non-interacting limit [$\ln(k_Fa_{2D}) =
  545$]. Since all effects considered in the present paper were too
weak to explain this damping, it must come from a different mechanism
which has not yet been identified.
\section*{Acknowledgements}
We thank E. Vogt for discussions. S.C. is supported by the {\it
  Funda\c{c}\~ao para a Ci\^encia e a Tecnologia} (FCT, Portugal) and
the {\it European Social Fund} (ESF) via the post-doctoral grant
SFRH/BPD/64405/2009.
\appendix
\section{Response function and determination of frequency and damping rate}
\begin{figure}
  \includegraphics[width=6cm]{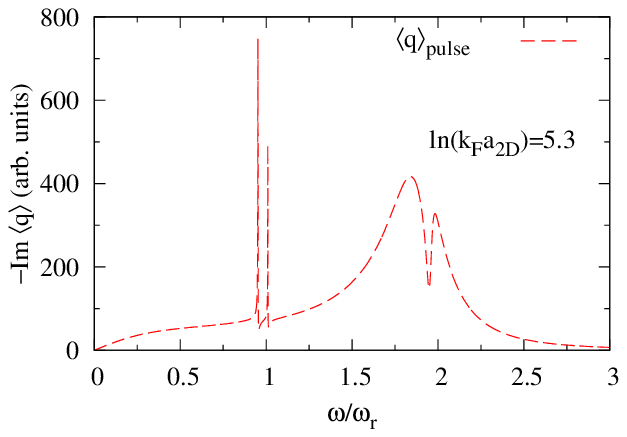}
  \includegraphics[width=6cm]{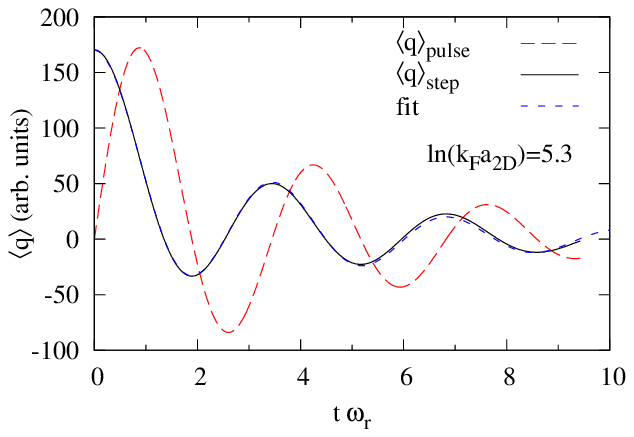}
  \caption{(Color online) Example for response functions in the frequency 
    (top) and time (bottom) domains. The upper panel shows $-\Im
    \ave{q}_\pulse(\omega)$ for the case $N = 4300$, $T/T_F = 0.47$,
    $\ln(k_Fa_{2D}) = 5.3$ in the trap potential shown in
    \Fig{fig:vtrap}. The lower panel shows the Fourier transform
    $\ave{q}_\pulse(t)$ (long dashes), the corresponding
    $\ave{q}_\step(t)$ (solid line) and the fit (short dashes) that
    yields the results $\omega = 1.876\omega_\rad$ and $\Gamma =
    0.234\omega_\rad$.
  \label{fig:QomegaQt}}
\end{figure}
As it was explained in \cite{Chiacchiera2011} and briefly mentioned in
\Sec{sec:moments}, the moments method gives at higher order a number
of complex eigenvalues $\omega_i$ whose real and imaginary parts
cannot directly be interpreted as frequencies and damping rates of
different collective modes. One rather has to look at the total
response function. The response to the $\delta$-pulse perturbation
\Eq{eq:Vpulse} can be written in the form
\begin{equation}
\ave{q}_\pulse(\omega) = \sum_{j=1}^n \frac{Z_j}{\omega-\omega_j}\,.
\end{equation}
This is Eq.~(25) of \Ref{Chiacchiera2011} if one replaces
$\omega_j-i\Gamma_j$ by a complex frequency $\omega_j$. The complex
frequencies $\omega_j$ satisfy $\Im \omega_j < 0$ (in the case of a
real $\omega_j$ one has to add an infinitesimal negative imaginary
part). A Fourier transform gives
\begin{equation}
\ave{q}_\pulse(t) = i \sum_{j=1}^n Z_j\, e^{-i\omega_j t}\, \theta(t).
\end{equation}
The response to a more realistic excitation which is adiabatically
switched on at $t = -\infty$ and which is suddenly switched off at $t =
0$ is given by
\begin{equation}
\ave{q}_\step(t) = \int_{-\infty}^0 dt' \ave{q}_\pulse(t-t')
  = \sum_{j=1}^n \frac{Z_j}{\omega_j} e^{-i\omega_j t\, \theta(t)}\,.
\end{equation}

Figure~\ref{fig:QomegaQt} shows a typical example for a response
function in the frequency and time domains. As a function of $\omega$
(upper panel), the response $-\Im \ave{q}_\pulse(\omega)$ has a couple
of spikes near $\omega=1$ coming from the (weak) coupling between
quadrupole and sloshing modes due to the asymmetry of the trap
potential. The broad peak corresponding to the quadrupole mode has a
sharp minimum near $\omega=2$ due to the interference between the
contributions of two complex eigenvalues. There is no obvious
prescription how to extract a unique $\omega$ and $\Gamma$ from this
response, so we transform the response to the time domain (lower
panel) and follow the method used in the analysis of the experiment
\cite{Vogt2012}, i.e., we fit \Eq{eq:Qfit} (short dashes) to
$\ave{q}_\step(t)$ (solid line) on the interval between $t = 0$ and $t
= 12$ ms $\approx 10/\omega_\rad$ \cite{Vogt_Thesis}.


\end{document}